\def\gax{\mathrel{\raise.3ex\hbox{$>$}\mkern-14mu\lower0.6ex\hbox{$\sim$}}}
\def\lax{\mathrel{\raise.3ex\hbox{$<$}\mkern-14mu\lower0.6ex\hbox{$\sim$}}}
\def\etal{{\it~et~al. }}
\def\eg{{\it e.g.:~}}
\def\angstrom{{\rm\,\AA}}

\def\mJy{{\rm\,mJy}}
\def\mJyBeam{{\rm\,mJy\,beam^{-1}}}
\def\kms{{\rm\,km\,s^{-1}}}
\def\kmsMpc{{\rm\,km\,s^{-1}\,Mpc^{-1}}}
\def\magsec{{\rm\,mag\,arcsec^{-2}}}

\documentstyle[11pt,aaspp4,flushrt]{article}
\begin{document}

\title{The Gravitationally-Lensed Radio Source MG\,0751+2716}

\centerline{\it Submitted to AJ: 1997.02.19\,. Accepted: 1997.04.15\,.}

\author{   J.~Leh\'ar\altaffilmark{1}, 
           B.F.~Burke\altaffilmark{2},
           S.R.~Conner\altaffilmark{2},
           E.E.~Falco\altaffilmark{1},
           A.B.~Fletcher\altaffilmark{2},
           M.~Irwin\altaffilmark{3},        
           R.G.~McMahon\altaffilmark{4}, 
           T.W.B.~Muxlow\altaffilmark{5},
           P.L.~Schechter\altaffilmark{2}
       }

\altaffiltext{1}{ Center for Astrophysics,
                  60 Garden St, Cambridge, MA 02138, USA.
                  jlehar@cfa.harvard.edu }
\altaffiltext{2}{ Mass. Inst. of Technology, Dept of Physics,
                  77 Mass. Ave, Cambridge, MA 02139, USA }
\altaffiltext{3}{ Royal Greenwich Observatory, 
                  Madingley Rd, Cambridge, CB3 0EZ, UK }
\altaffiltext{4}{ Institute of Astronomy,  
                  Madingley Rd, Cambridge, CB3 0HA, UK }
\altaffiltext{5}{ Nuffield Radio Astronomy Laboratories, 
                  Jodrell Bank, Cheshire SK11 9DL, UK }

\begin{abstract}

\noindent
We report the discovery of a new gravitationally lensed radio source. 
Radio maps of MG\,0751+2716 show four lensed images,
which, at higher resolution, are resolved into long arcs of emission.
A group of galaxies is present in optical images, 
including the principal lensing galaxy,
with a much brighter galaxy just a few arcseconds away. 
We have measured the redshift of this brighter galaxy. 
No optical counterpart to the background source has been detected.
Lens models that can readily reproduce the lensed image positions
all require a substantial shear component. 
However, neither the very elongated lens nor the bright nearby galaxy 
are correctly positioned to explain the shear. 
Lens models which associate the mass with 
the light of galaxies in the group
can produce an acceptable fit, but only with an extreme 
mass-to-light ratio in one of the minor group members. 

\end{abstract}


\clearpage

\section{Introduction}
\nopagebreak[4]

The past decade has seen great progress 
in the study of gravitational lensing (\eg\cite{sch92}; \cite{bla92}). 
Since extragalactic radio sources are typically at high redshift,
radio surveys have proven to be very effective for finding lensed systems. 
The MG-VLA (\cite{law86}; \cite{hew86}; \cite{leh91}; \cite{her96}), 
JVAS (\cite{pat92}), and CLASS (\cite{mye95}) radio searches
have between them found about half of the $\sim20$ confirmed cases, 
at a rate of approximately one lensed system per thousand sources. 
Most of these systems have extended structures in the radio
which can provide many lens model constraints. 

Here, we present the discovery of a sixth lensed system 
from the MG-VLA search, and later observations
which support the lensing interpretation. 
We also use the radio components to determine lens models,
and compare the lensing mass to the observed luminosity 
distribution of the foreground galaxies.

\section{Observations}
\nopagebreak[4]

MG\,0751+2716 was discovered as part of the MG-VLA search, 
and a number of NRAO Very Large Array (VLA) observations 
were acquired shortly thereafter (see Table~1).
The VLA observations were performed in standard continuum mode, 
interleaved with observations of J0746+25, 3C\,286, and OQ\,208, 
to calibrate the antenna phases, flux scales and polarizations, respectively.
Calibration and mapping were performed using standard VLA procedures. 
The maps were further self-calibrated twice, 
except for the 2\,cm map, which had insufficient dynamic range.
The resulting maps have rms noise levels of about twice the thermal limit. 
Figure~1 shows the VLA maps, with the components named in order of
decreasing 3.6\,cm peak flux density. 
There are four components, separated by $\approx0\farcs8$, 
with a morphology similar to that of 
the lensed system MG\,0414+0534 (\cite{hew92}). 
Polarized emission was only detected at 3.6\,cm,
where all four components show 
the same linearly polarized fraction and orientation. 
Slight flux density variations are seen 
between the two 3.6\,cm observations
($\sim$5\% in the B~component),
but their significance is marginal ($\approx$2 standard deviations).
The radio spectra of the four components agree, 
and no other radio emission was detected nearby at any frequency. 
The A1 component on the 2.0\,cm map is located at
$\alpha${}={}07:51:41.49,
$\delta${}={}$+27$:16:31.6 (J2000),
with an uncertainty of $\sim0\farcs2$\ (\cite{law86}). 
The VLA component photometry (for rectangular apertures) is given in Table~2. 

We also observed MG\,0751+2716 with the MERLIN array, 
using six antennae 
(Cambridge, Darnhall, Defford, NRAL~Mk\,II, Knockin, Tabley). 
The observations were phase-referenced to 
the compact calibrator 0743+277, with a 10\,min cycle time. 
The strong point source 2134+004 was used to 
determine the bandpass and non-closing corrections,
and the flux calibration was determined using 3C\,286. 
The antenna polarizations were calibrated
using 0743+277 and 3C\,286. 
The data were edited, calibrated and mapped using 
standard MERLIN analysis procedures,
with several interations of self-calibration. 
The resulting map resolves the VLA components into long arcs,
with the A~and C~arcs merging. 
The arcs are unresolved radially, 
but significant maxima can be used to identify subcomponents
(see~Figure~2). 
The MERLIN A1~component is at 
$\alpha${}={}07:51:41.48,
$\delta${}={}$+27$:16:31.6 (J2000),
consistent VLA 2.0\,cm position
to within $\approx${}$0\farcs13$\,.
The positions and peak intensities of the subcomponents
are given in Table~3, relative to component~A1.
No polarized emission was detected in our MERLIN data. 

We used the Cambridge Automated Plate Measuring (APM) machine catalog
to locate optical counterparts on 
the Palomar Observatory Sky Survey exposures. 
No direct optical counterpart was found, but
a 19th magnitude galaxy, G1, 
is located $\sim6''$ West of the radio source, at
$\alpha${}={}07:51:41.04,
$\delta${}={}$+27$:16:32.6 (J2000). 
The astrometric uncertainty of the APM catalog is $\sim0\farcs5$. 

We obtained an optical spectrum of G1, 
using the ISIS spectrograph 
on the William Herschel Telescope at La~Palma. 
We also observed a spectral flux calibration star G191$-$B2B (\cite{mas88}).
The data were reduced following standard procedures 
in the IRAF software package, 
and the resulting spectrum is shown in Figure~3. 
We identified several absorption features common to old stellar populations, 
yielding a redshift of $z=0.351$ with an uncertainty of $\sim0.001$. 
The Mg\,b absorption is resolved into two lines, 
separated by $\sim14\angstrom$, suggesting an upper limit
on the stellar velocity dispersion of $\sigma_v\lax250\kms$. 

A number of optical CCD images were also obtained for 
the MG\,0751+2716 field (see Table~1),
using the MDM Hiltner telescope, 
and the Isaac Newton Telescope (INT) at La~Palma. 
The data were reduced using standard procedures in IRAF. 
The INT image was taken under photometric conditions
and calibrated using the SA\,92 field of standard stars (\cite{lan92}).
The MDM $R$~filter image was then calibrated to the INT observation
using a star (Star\,A) near G1. 
Figure~4 shows the optical field near MG\,0751+2716,
with some of the galaxies named in order of increasing distance from G1.
We used FOCAS (\cite{val82}) to obtain
positions and magnitudes for all the objects on the MDM images, 
using 5\,pixel radius apertures, and limiting magnitudes of
$R${}$\approx${}25 and $I${}$\approx${}23.5\,.
The $R${}$-${}$I$ colors are scaled to G1, whose absolute color is unknown. 
Assuming a flat optical $\nu{}f_\nu$ spectrum, however,
G1 would have $R${}$-${}$I\approx0.47$\,.
The object positions were calibrated to the APM Palomar data, 
using twelve stellar objects in the field. 
Table~4 gives the results for the most relevant objects. 
A complete listing can be obtained from the authors by request. 
G3 lies $0\farcs2$ Northeast of the MERLIN~A1 component, 
coincident within the APM positional uncertainty, 
and is probably the principal lensing galaxy. 

We fitted surface brightness profiles to G1 and G3, 
using the $R$~filter MDM image. 
Elliptical profiles were convolved in two dimensions with
the observed Star\,A point spread function. 
A goodness-of-fit $\chi^2$ was determined by summing 
the squared differences between the convolved model 
and the observed data for each pixel,
normalized to the off-source sky rms. 
Three models were considered:
a DeVaucouleurs $r^{1/4}$ profile, 
where $R_0$ and $I_0$ refer to the effective radius; 
an exponential profile; and a Hubble profile. 
In each case we varied the profile shape, 
the galaxy position and ellipticity, and the sky level.
The best fit was found using an ``amoeba'' search (\cite{pre89}).
We estimated the parameter uncertainties 
from the $\Delta\chi^2${}$<${}$1$ intervals,
by stepping each parameter away from the best fit, 
while optimizing the others.
The results are shown in Table~5.  
The G1 models have excess residuals ($\sim$10\% of the peak intensity)
near the center of the galaxy, and the exponential model is excluded. 
The profile parameters $R_0$ and $I_0$ are very sensitive
to problems with the deconvolution, 
but the models all give similar ellipticities and orientations. 
For G3, the models could not be distinguished, 
but all had consistent ellipticities and orientations.

\section{Discussion}
\nopagebreak[4]

MG\,0751+2716 is almost certainly gravitationally lensed. 
The radio morphology is very characteristic of gravitational lensing, 
and extremely unusual for any other interpretation. 
As expected for lensed images, the four components have 
matching radio spectra and polarizations, 
and the 2.0\,cm VLA and 6.0\,cm MERLIN maps are in good agreement. 
There is also a good optical candidate G3 for the lens galaxy, 
consistent with the radio source position. 
The lens redshift is not yet determined, 
but its color, and proximity to G1, suggest that 
G3 is associated with G1, 
at a redshift of $z${}$=${}0.351\,.

There is no optical counterpart to the background radio source. 
Although it is not unusual for lensed radio sources
to have very faint host galaxies, any evidence of the source
would confirm the lensing and reduce the model uncertainties. 
Since the radio source has a steep spectrum and is extended,
it could be close to the core of its host galaxy, 
or off to the side in a radio lobe. 
In the former case, 
the host galaxy should be visible as a faint ring of arcs around the lens.
The present observations, however, lack the sensitivity and angular resolution
to detect such a ring.
It is unlikely that the host galaxy is off to the side, 
since the deep 20\,cm VLA observation shows
no clear evidence of a nearby core or another radio lobe within $1'$. 
When the 20\,cm VLA map is super-resolved 
with a circular $1\arcsec$ restoring beam, 
a $\sim2\mJy$ component does appear 
about $2\arcsec$ East of the radio source
(along the major axis of the natural radio beam),
which coincides with an $R\sim25$ residual in the
optical profile fits to G1;
but both the radio and optical detections of this object
are marginal at best. 

We used the MERLIN components (Table~3) to constrain
models of the lensing mass distributions. 
The lensed image subcomponents are labelled to indicate which 
are likely to have common background source components (see~Figure~2). 
We used the ``lensmod'' software (\cite{leh93}), 
modified to operate in the image plane. 
For a given lens model and source component, 
the fitting algorithm traces the observed images to the source plane, 
and then finds the model image positions which best correspond 
to the average source component. 
A $\chi^2$ goodness-of-fit is then computed from
the differences between the observed and model image positions. 
We used the ``amoeba'' algorithm (\cite{pre89})
to find the parameters with the minimum $\chi^2$. 

As a first approximation, 
three simple mass models were considered:
P+$\gamma$, a point mass lens with an external shear; 
SIP+$\gamma$, a singular isothermal potential with an external shear; 
and ESIP, an elliptical singular isothermal potential. 
These models have five parameters: 
the lens coordinates $x_{_L}$,$y_{_L}$; 
the ring size $\beta$; 
either an external shear $\gamma$ or an ellipticity $\epsilon$;
and the major axis orientation angle, $\phi_{\gamma}$ or $\phi_{\epsilon}$.
The parameter $\epsilon$ approximates the $(1-b/a)$ isodensity ellipticity,
derived from the isopotential eccentricity $\epsilon_{_{BK}}$
(\cite{bla87}).
The best fit parameters are given in Table~6, 
with uncertainties derived from the $\Delta\chi^2${}$<${}$1$ ranges. 
None of these simple models provides 
a satisfactory fit to the observed images.
Moreover, the orientations of the best fit lens models are consistent with
neither the ellipticity of G3 nor the location of G1 relative to G3. 

Because our single-mass models failed, 
we explored models which approximate the effects of the nearby galaxies:
G3+SIP, an ESIP with its orientation constrained 
to have $\phi_{\epsilon}=\phi_{_{G3}}$ (see~Table~5),
and an additional SIP which can be freely adjusted and moved to a position
specified by the polar coordinates $r_{ext}$ and $\phi_{ext}$
(with an effective shear of $\approx{}\beta_{ext}/r_{ext}$);
G3+Grp, an ESIP constrained to $\phi_{_{G3}}$,
with four external SIPs, at the positions of the other galaxies in the group,
and with ring sizes $\beta_{Gi}$ proportional to the square-root of their 
optical brightnesses (see Table~4), scaled to $\beta_{G1}$
(assuming a common redshift of $z_{_{G1}}$ for all the group galaxies); 
and G3+4G, like G3+Grp, 
but with independently adjustable group member ring sizes. 
The G3+SIP model provides a good fit, and the G3+4G does not. 
The G3+4G model can account for the shear by moving mass to G4. 
Figure~5 gives a graphic illustration of the fit quality. 
For all of our models, a common linear arrangement of the 
source components can reproduce the observed image structures. 
Assuming $z_{_{G3}}=0.351$,
$H_0=75\kmsMpc$, $\Omega=1$, 
and a flat optical spectral energy distribution 
(constant $\nu{}f_\nu$, or $B${}$-${}$R\approx1$) with no evolution, 
the Faber-Jackson relation (as~in~\cite{fuk91})
implies a velocity dispersion of
$\approx140\kms$ for G3. 
This is consistent with 
an isothermal ring size of $\beta_{_{G3}}=0\farcs37\pm0\farcs02$,
provided that $z_{_S}\gax1$\,.
Likewise, G1 has a Faber-Jackson velocity dispersion estimate of 
$\approx240\kms$. 
This dispersion is consistent with $\beta_{_{G1}}=0\farcs8\pm0\farcs1$
for $z_{_S}\gax0.8$, and with the upper limit from the optical spectrum. 

MG\,0751+2716 appears to be one of the increasing number of lensed systems
which require more shear than can be easily accounted for optically
(\cite{kee97}). 
The principal shortcoming of the G3+SIP model 
is that it places the external mass 
seven standard deviations South of G1. 
Similarly, the G3+4G model can only account for the image positions
if $\beta_{_{G4}}>\beta_{_{G1}}$, 
despite the fact that G4 is ten times fainter than G1. 
This peculiar result cannot be easily explained by 
placing G4 at a different redshift. 
We computed the expected ring size 
for a galaxy with an observed magnitude at arbitrary $z$,
assuming a non-evolving flat spectral energy distribution.
For a given observed optical magnitude and assumed M/L ratio,
a faint lens needs to be more distant, and thus more luminous, 
to produce a large ring size. 
We found that for G4 and G3 to have the same M/L ratio, 
the lens would have to be at $z\gax30$
with a background radio source at $z>100$, 
for any $\Lambda=0$ cosmology with $\Omega\gax0.2$\,.
Since G4 cannot reasonably account for the required shear,
either there must be a dark concentration 
of matter somewhere to the South of G1, 
or G3 must be embedded in a dark halo
whose orientation is misaligned with G3 by $>${}$30^{\circ}$. 

Considerable work remains to be done on this system. 
The distance to the lens G3 is not known, 
and the mass estimates depend sensitively on the lens redshift. 
Nor are the distances known to most of the other galaxies in the group.
The source redshift is also unknown, of course,
and although the model results are not very sensitive to $z_{_S}$,
any optical evidence of the background
source would be an important confirmation. 
HST observations are planned that may be able to distinguish
the two possibilities for the background source location. 
We also plan to continue conducting spectroscopic observations of this system
to determine redshifts and velocity dispersions 
for G3 and the other group members. 
All of our models predict time delays 
of a few hours between lensed images. 
Since the radio source is extended, has a steep spectrum, 
and has not varied strongly over a two-year period
it is unlikely that delays can be measured for MG\,0751+2716. 
However, the radio spectrum does flatten 
towards the West end of the B~component,
thus offering some hope that there is a compact, variable core. 
We have obtained deep 2\,cm VLA observations which should help determine
whether such a core is present, and provide more detail on the lensed images. 
MG\,0751+2716 is well suited for the ``lensclean'' algorithm
(\cite{koc93}), 
in which a model of the background source is constructed
to be consistent with the observed extended structures. 
With C.~Kochanek, we have started this analysis,
and we hope to obtain more sensitive constraints 
on the lensing mass distribution. 

\acknowledgments
It is a pleasure to thank
Chris Kochanek and Chuck Keeton for help with the lens models,
and Karl Glazebrook for help with the optical spectrum. 
The NRAO is operated by Associated Universities, Inc.,
under cooperative agreement with the National Science Foundation.
Observations reported here were obtained, in part, 
at the MDM Observatory;
a consortium of the University of Michigan, Dartmouth College
and the Massachusetts Institute of Technology.
IRAF is distributed by the National Optical Astronomy
Observatories, which are operated by the Association of Universities for
Research in Astronomy, Inc., under cooperative agreement with the National
Science Foundation.
JL gratefully acknowledges support from NSF grant AST93-03527.



\begin{deluxetable}{llllll}
\tablecolumns{6}
\tablecaption{Observation Log}
\tablehead{
   \colhead{Date}      &
   \colhead{Telescope}        &
   \colhead{Wavelength}  &
   \colhead{Exposure}       &
   \colhead{Resol (fwhm)}          &
   \colhead{Noise (rms)}           
}
\startdata
\sidehead{Radio Observations}
1990.05.03 & VLA (A)   & 3.6\,cm &
    2.5\,min & $0\farcs2$   & $0.29\mJyBeam$ \\
1992.11.18 & VLA (A)   & 2.0\,cm &
    5\,min   & $0\farcs11$  & $0.53\mJyBeam$ \\
{}{}{}     &           & 3.6\,cm &
    2.5\,min & $0\farcs20$  & $0.18\mJyBeam$ \\
{}{}{}     &           & 6.2\,cm &
    1.5\,min & $0\farcs33$  & $0.28\mJyBeam$ \\
1993.02.03 & VLA (BnA) & 20.4\,cm  &
    15\,min  & $4\farcs2\times1\farcs4$ & $0.14\mJyBeam$ \\
1992.11.25 & MERLIN    & 6.0\,cm &
    15.5\,hr & $0\farcs05$  & $0.089\mJyBeam$ \\
\sidehead{Optical Spectroscopy}
1992.04.03 & WHT 4\,m  & $6992\pm${}$1800\angstrom$ & 
    60\,min  & $2.5\angstrom/$pixel & $\sim${}5\,$\mu$Jy/pixel \\
\sidehead{Optical Imaging}
1993.11.19 & MDM 2.4\,m      & $I$~filter & 
    10\,min  & $0\farcs8$   & $\sim24.6\magsec$ \\
1994.01.10 & INT 2\,m        & $R$~filter & 
    15\,min  & $1\farcs5$   & $25.6\magsec$ \\
1995.12.18 & MDM 2.4\,m      & $R$~filter & 
    1.8\,hr  & $0\farcs8$   & $26.2\magsec$ \\
\enddata
\end{deluxetable}

\begin{deluxetable}{lrrrr}
\tablecolumns{5}
\tablecaption{VLA Photometry}
\tablehead{
   \colhead{ Component                          } &
   \colhead{ Wavelength                         } &
   \colhead{ $S_{aper}$ (mJy)                   } &
   \colhead{ $S_{pol}$\tablenotemark{a} (\%)    } &
   \colhead{ $\phi_{pol}$\tablenotemark{a}      }     
}
\startdata
   A+C              &
   $2.0$\,cm            &
   $31.11\pm2.05$   &
   $0.0\pm6.6$      &
   \nodata          \\
                    &
   $3.6$\,cm\tablenotemark{b}     &
   $61.36\pm0.90$   &
   $7.7\pm1.2$      &
   $0$              \\
                    &
   $3.6$\,cm            &
   $62.24\pm0.49$   &
   $7.6\pm0.8$      &
   $+3$             \\
                    &
   $6.2$\,cm            &
   $118.69\pm0.61$  &
   $0.0\pm0.5$      &
   \nodata          \\
\\
   B                &
   $2.0$\,cm            &
   $14.28\pm1.25$   &
   $0.0\pm8.8$      &
   \nodata          \\
                    &
   $3.6$\,cm\tablenotemark{b}          &
   $23.48\pm0.65$   &
   $12.0\pm2.1$     &
   $+4$             \\
                    &
   $3.6$\,cm            &
   $24.75\pm0.32$   &
   $9.0\pm1.3$      &
   $+7$             \\
                    &
   $6.2$\,cm            &
   $44.09\pm0.42$   &
   $0.0\pm1.0$      &
   \nodata          \\
\\
   D                &
   $2.0$\,cm            &
   $5.73\pm0.91$    &
   $0.0\pm16.0$     &
   \nodata          \\
                    &
   $3.6$\,cm\tablenotemark{b}          &
   $11.45\pm0.49$   &
   $15.5\pm4.4$     &
   $-6$             \\
                    &
   $3.6$\,cm            &
   $12.18\pm0.30$   &
   $9.4\pm2.6$      &
   $+3$             \\
                    &
   $6.2$\,cm            &
   $17.53\pm0.36$   &
   $0.0\pm2.0$      &
   \nodata          \\
\\
   TOTAL            &
   $2.0$\,cm            &
   $47.75\pm4.22$   &
   $0.0\pm8.8$      &
   \nodata          \\
                    &
   $3.6$\,cm\tablenotemark{b}          &
   $103.71\pm1.52$  &
   $13.7\pm1.3$     &
   \nodata          \\
                    &
   $3.6$\,cm            &
   $104.60\pm0.88$  &
   $11.7\pm1.2$     &
   \nodata          \\
                    &
   $6.2$\,cm            &
   $191.16\pm1.02$  &
   $0.0\pm0.5$      &
   \nodata          \\
                    &
   $20.4$\,cm           &
   $412.99\pm0.42$  &
   \nodata          &
   \nodata          \\
\enddata
\tablenotetext{a}{ Linear polarization fractional intensity 
                   and orientation in degrees from North through East. }
\tablenotetext{b}{ From the 1991.03.05 MG-VLA observation. }
\end{deluxetable}

\begin{deluxetable}{lrrrl}
\tablecolumns{5}
\tablecaption{MERLIN Components}
\tablehead{
   \colhead{ Component                       } &
   \colhead{ $\Delta\alpha$\tablenotemark{a} } &
   \colhead{ $\Delta\delta$\tablenotemark{a} } &
   \colhead{ $S_{peak}$\tablenotemark{b}     } &
   \colhead{ Source ID\tablenotemark{c}      }        
}
\startdata
   A1               &
   $0.000$          &
   $0.000$          &
   $24.603$         &
   1                \\
   A2(?)            &
   $-0.037$         &
   $+0.045$         &
   $9.265$          &
   2 (unused, for ref only)    \\
   A3               &
   $-0.060$         &
   $+0.127$         &
   $4.652$          &
   3          \\
   A4               &
   $-0.067$         &
   $+0.180$         &
   $5.086$          &
   4          \\
\\
   B1               &
   $+0.495$         &
   $-0.270$         &
   $8.470$          &
   1          \\
   B2               &
   $+0.405$         &
   $-0.285$         &
   $7.888$          &
   2          \\
   B3$-$B4(?)         &
   $+0.315$         &
   $-0.292$         &
   $6.127$          &
   3,4 (uncertain) \\
   B5               &
   $+0.278$         &
   $-0.292$         &
   $9.266$          &
   5          \\
   B6               &
   $+0.067$         &
   $-0.308$         &
   $0.889$          &
   6          \\
\\
   C1               &
   $+0.082$         &
   $+0.510$         &
   $8.971$          &
   1          \\
   C2               &
   $+0.022$         &
   $+0.450$         &
   $9.021$          &
   2          \\
   C3               &
   $-0.052$         &
   $+0.300$         &
   $5.414$          &
   3          \\
   C4               &
   $-0.060$         &
   $+0.255$         &
   $5.020$          &
   4          \\
\\
   D1$-$D5 &
   $+0.637$\tablenotemark{d}         &
   $+0.345$\tablenotemark{d}         &
   $15.315$         &
   1,2,3,4,5        \\
\enddata
\tablenotetext{a}{ Offsets are in arcsec from A1, 
                   with 8\,mas (1\,pixel) assumed uncertainties. }
\tablenotetext{b}{ Peak flux densities in $\mJyBeam$, not used for model fits. }
\tablenotetext{c}{ Source identifications (see Figure~2);
                   multiple numbers imply blended components. }
\tablenotetext{d}{ Assumed a $16\times8$\,mas error ellipse, 
                   oriented at $+45\deg$ from North through East. }
\end{deluxetable}

\begin{deluxetable}{lrrrrr}
\tablecolumns{6}
\tablecaption{Optical Components}
\tablehead{
   \colhead{Object}           &
   \colhead{$\Delta\alpha$\tablenotemark{a} }   &
   \colhead{$\Delta\delta$\tablenotemark{a} }   &
   \colhead{$R_{is}$\tablenotemark{b} }         &
   \colhead{$R_{ap}$\tablenotemark{c} }         &
   \colhead{$R${}$-${}$I$\tablenotemark{c} }    
}
\startdata
G1         &
   $0.00$     &
   $0.00$     &
   $19.09$    &
   $19.98$    &
   $0.00$     \\
G2         &
   $+4.07$    &
   $-3.59$    &
   $23.25$    &
   $22.90$    &
   \nodata    \\
G3         &
   $+5.83$    &
   $-0.91$    &
   $21.31$    &
   $21.36$    &
   $-0.04$    \\
G4         &
   $-0.91$    &
   $-6.81$    &
   $22.23$    &
   $22.24$    &
   $-0.38$    \\
G5         &
   $+10.29$   &
   $-3.64$    &
   $22.30$    &
   $22.27$    &
   \nodata    \\
Star\,A    &
   $+10.40$   &
   $+14.78$   &
   $19.00$    &
   $19.09$    &
   $0.11$     \\
\enddata
\tablenotetext{a}{ Offsets from G1 are in arcsec. }
\tablenotetext{b}{ Magnitudes for light enclosed by 
                   the $R=25\magsec$ isophote. }
\tablenotetext{c}{ Aperture magnitudes within a $1\farcs4$ radius. }
\end{deluxetable}

\begin{deluxetable}{lrrrrrr}
\tablecolumns{7}
\tablecaption{Optical Profile Models}
\tablehead{
   \colhead{~}                &
   \multicolumn{3}{c}{G1}     &
   \multicolumn{3}{c}{G3}     \\
   \cline{2-4}                
   \cline{5-7}                
   \colhead{Model}            &
   \colhead{DeVauc}           &
   \colhead{Expon}            &
   \colhead{Hubble}           &
   \colhead{DeVauc}           &
   \colhead{Expon}            &
   \colhead{Hubble}           
}
\startdata
   NDOF       &
   7088       &
   7088       &
   7088       &
   229        &
   229        &
   229        \\
   $\chi^2/DOF$ &
   2.30         &
   3.64         &
   1.86         &
   1.13         &
   1.17         &
   1.37         \\
   $\Delta\alpha$    &
   $-0.114\pm0.006$  &
   $-0.109$          &
   $-0.110$          &
   $+0.011\pm0.003$  &
   $+0.011$          &
   $+0.007$          \\
   $\Delta\delta$    &
   $+0.107\pm0.004$  &
   $+0.095$          &
   $+0.098$          &
   $+0.024\pm0.003$  &
   $+0.030$          &
   $+0.024$          \\
   $\epsilon$        &
   $0.417\pm0.003$   &
   $0.321$           &
   $0.398$           &
   $0.49\pm0.03$     &
   $0.48$            &
   $0.45$            \\
   $\phi_\epsilon$   &
   $+1.5\pm0.4$     &
   $-5.0$           &
   $+1.4$           &
   $+16.7\pm2.0$    &
   $+14.9$          &
   $+15.9$          \\
   $R_0$             &
   $1.98\pm0.02$     &
   $0.78$            &
   $0.48$            &
   $0.27\pm0.03$     &
   $0.29$            &
   $0.0097$          \\
   $I_0$             &
   $22.77\pm0.02$    &
   $20.12$           &
   $19.75$           &
   $20.32\pm0.37$    &
   $19.70$           &
   $13.23$           \\
\enddata
\tablenotetext{}{
  Profile center offsets are in arcsec
  from center of the brightest pixel.
  Profile ellipticity $\epsilon=1${}$-${}$b/a$,
  and major axis orientation $\phi_\epsilon$
  is in degrees from North through East.
  $R_0$~is the scale radius in arcsec,
  and $I_0$~is the surface brightness scale
  expressed in $R\magsec$.
  Uncertainties, from the $\Delta\chi^2${}$<${}$1$ ranges, 
  are given for the DeVaucouleurs model;
  those for other models are comparable.
}
\end{deluxetable}

\begin{deluxetable}{lrrrrrr}
\tablecolumns{7}
\tablecaption{Lens Model Results}
\tablehead{
   \colhead{Model}               &
   \colhead{P+$\gamma$}          &      
   \colhead{SIP+$\gamma$}        &      
   \colhead{ESIP}                &      
   \colhead{G3+SIP}              &      
   \colhead{G3+Grp}              &      
   \colhead{G3+4G}                      
}
\startdata
   NDOF       &
   15         &
   15         &
   15         &
   13         &
   15         &
   12         \\
   $\chi^2/DOF$ &
   1.78         &
   1.96         &
   3.95         &
   1.10         &
   3.64         &
   1.14         \\
   $\Delta\alpha$   &
   $+0.346$&
   $+0.348$&
   $+0.356$&
   $+0.344\pm0.005$ &
   $+0.336\pm0.004$ &
   $+0.345\pm0.005$ \\
   $\Delta\delta$   &
   $+0.164$&
   $+0.164$&
   $+0.176$&
   $+0.160\pm0.005$ &
   $+0.158\pm0.004$ &
   $+0.163\pm0.004$ \\
   $\beta$          &
   $0.398$&
   $0.402$&
   $0.408$&
   $0.371\pm0.004$  &
   $0.351\pm0.005$  &
   $0.352\pm0.005$  \\
   $\gamma$ or $\epsilon$ &
   $0.188$&
   $0.093$&
   $0.344$&
   $0.092\pm0.024$  &
   $0.224\pm0.018$  &
   $0.125\pm0.026$  \\
   $\phi_{\gamma}$ or $\phi_{\epsilon}$ &
   $ -116.6$&
   $ -116.5$&
   $ -116.2$&
   $ -163.6\pm3.3$   &
   $ -155.5\pm2.6$   &
   $ -164.0\pm3.3$   \\
   $\beta_{ext}$    &
   \nodata          &
   \nodata          &
   \nodata          &
   $0.63^{+0.5}_{-0.2}$ &
   \nodata          &
   \nodata          \\
   $r_{ext}$        &
   \nodata          &
   \nodata          &
   \nodata          &
   $4.0^{+2.5}_{-1.0}$ &
   \nodata          &
   \nodata          \\
   $\phi_{ext}$     &
   \nodata          &
   \nodata          &
   \nodata          &
   $-106.9\pm2.7$    &
   \nodata          &
   \nodata          \\
   $\beta_{_{G1}}$  &
   \nodata          &
   \nodata          &
   \nodata          &
   \nodata          &
   $0.94\pm0.06$    &
   $0.76\pm0.08$    \\
   $\beta_{_{G2}}$  &
   \nodata          &
   \nodata          &
   \nodata          &
   \nodata          &
   ($0.060$)        &
   $<0.04$          \\
   $\beta_{_{G4}}$  &
   \nodata          &
   \nodata          &
   \nodata          &
   \nodata          &
   ($0.222$)        &
   $1.11\pm0.13$    \\
   $\beta_{_{G5}}$  &
   \nodata          &
   \nodata          &
   \nodata          &
   \nodata          &
   ($0.215$)        &
   $<0.04$          \\
\enddata
\tablenotetext{}{
  Lens model center offsets are in arcsec from A1,
  and ring sizes $\beta$, are in arcsec.
  Orientation angles are in degrees from North through East.
  External shears stretch images perpendicular to $\phi_\gamma$,
  which aligns with the tidal mass concentration. 
  For the G3+Grp model, the numbers in parentheses show
  the other ring sizes, which are directly scaled to $\beta_{_{G1}}$.
  Uncertainties are from the $\Delta\chi^2${}$<${}$1$ ranges. 
}
\end{deluxetable}



\begin{figure} 
\epsscale{0.9}
\plotone{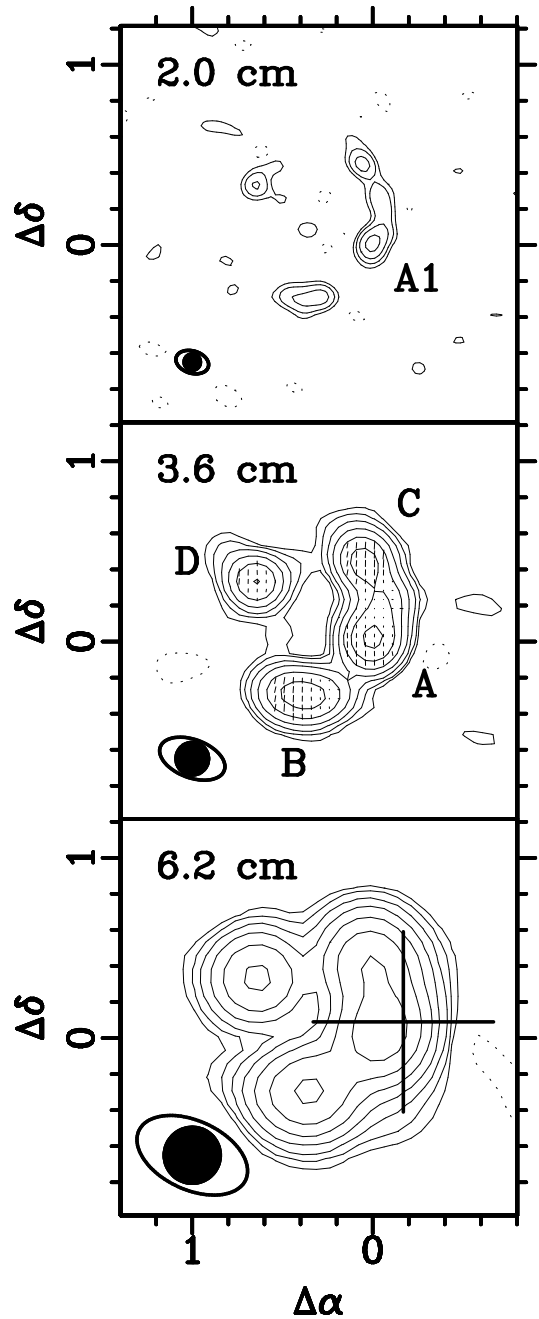}
\caption
{
VLA maps of MG0751+2716.
Offsets are in arcseconds relative to A1.
The contours increase by factors of two 
from $2\sigma_{rms}$, where $\sigma_{rms}$ is the off-source rms. 
Fractional polarization vectors are scaled so that one vector spacing
corresponds to 20\%, and are oriented along the electric field. 
Each panel shows the fwhm uniform-weighted (open) 
and restoring (filled) beams. 
The optical position of G3 is shown as a cross in the 6.2\,cm map, 
scaled to the $\pm0\farcs5$ APM astrometric uncertainty.
}
\end{figure}

\begin{figure}
\epsscale{0.9}
\plotone{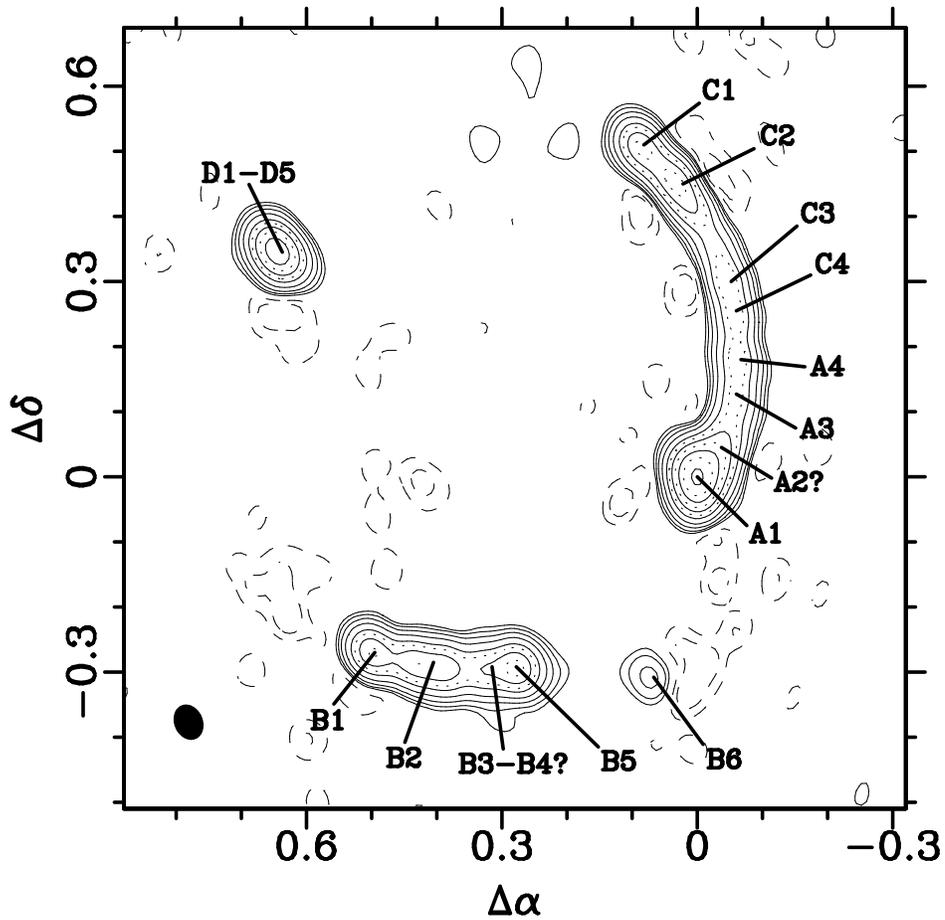}
\caption
{
MERLIN map of MG0751+2716.
Offsets are in arcseconds relative to A1.
The contours increase by factors of two from $2\sigma_{rms}$,
where $\sigma_{rms}$ is the off-source rms. 
Additional dotted contours increase by factors of $\sqrt{2}$
from $32\sigma_{rms}$ to emphasize the labelled subcomponents. 
The natural-weighted fwhm beam is shown at lower left.
}
\end{figure}

\begin{figure}
\epsscale{0.9}
\plotone{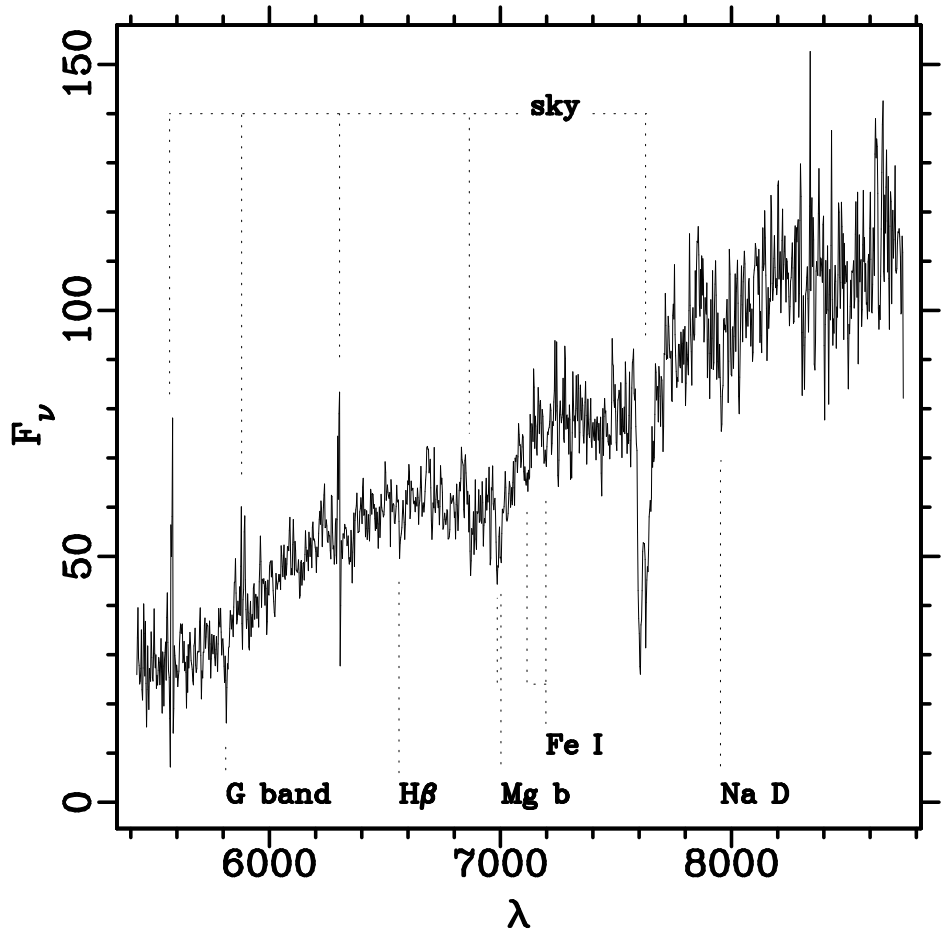}
\caption
{
Optical spectrum of G1. 
Wavelength $\lambda$ is in angstroms, 
and the flux density $f_\nu$ is in $\mu$Jy. 
Several absorption features are seen at $z=0.351$, 
and some prominent sky features are also labelled.
}
\end{figure}

\begin{figure}
\epsscale{0.9}
\plotone{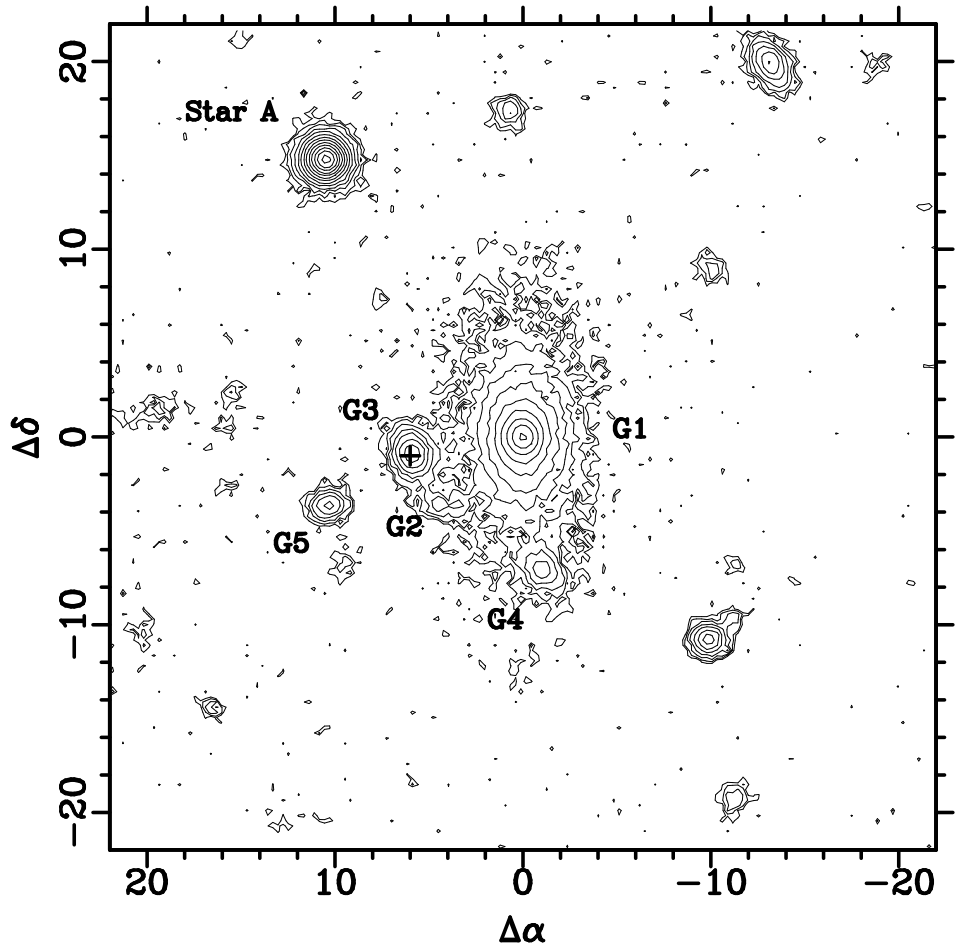}
\caption
{
Optical field around MG0751+2716, from the MDM $R$~filter image. 
The contours increase from $2\sigma_{rms}$
by factors of half a magnitude,
where $\sigma_{rms}$ is the off-source rms. 
Offsets are in arcsec from G1, 
and the position of the VLA 2\,cm component A1 is shown 
as a $\pm0\farcs5$ cross, reflecting the APM astrometric error.
}
\end{figure}

\begin{figure}
\epsscale{0.8}
\plotone{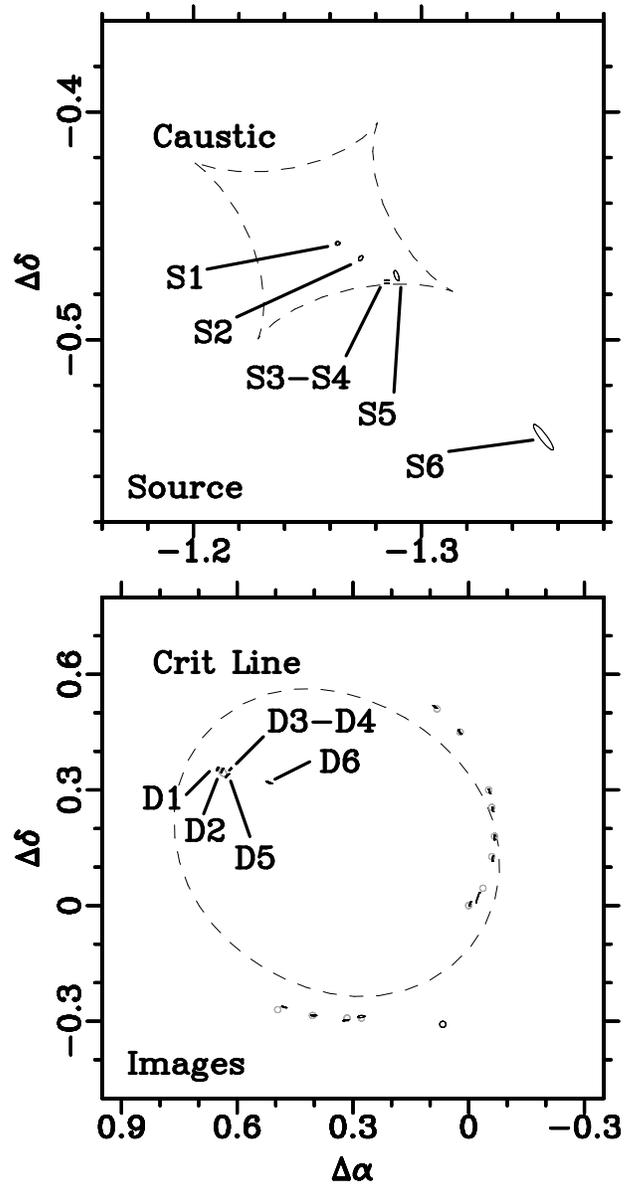}
\caption
{
Schematic view of the G3+4G lens model.
Offsets are given in arcsec from the MERLIN A1 component. 
The source plane shows the error ellipses for
the average source components. 
The image plane shows these average components
projected to the image plane (solid error ellipses),
as well as the observed image component positions (dotted error ellipses). 
The source and D~image components have been labelled for reference. 
Note that the source components are shifted substantially,
due to the influence of the group galaxies. 
}
\end{figure}

\end{document}